\documentclass[amsmath,amssymb,aps,longbibliography,nofootinbib,twocolumn,superscriptaddress]{revtex4-2}

\usepackage{graphicx}
\usepackage{dcolumn}
\usepackage{bm}
\usepackage{bbm}
\usepackage{bbold}

\usepackage{tikz}

\usetikzlibrary{quantikz}

\DeclareMathSymbol{\mathbbE}{\mathord}{AMSb}{"45}
\newcommand{\ex}{\mathbbE}

\newcommand{\psiT}{\Psi_{\textrm{T}}}
\newcommand{\psiiT}{\psi_i^{\textrm{T}}}
\newcommand{\psijT}{\psi_j^{\textrm{T}}}
\newcommand{\LUCJ}{\Psi_{\mathrm{LUCJ}}}
\newcommand{\EiL}{E_i^{\textrm{L}}}
\newcommand{\Hfn}{H^{\textrm{fn}}}
\newcommand{\Efn}{E^{\textrm{fn}}}
\newcommand{\Pfn}{P^{\textrm{fn}}}
\newcommand{\EvarLUCJ}{E_{\mathrm{var}}^{\mathrm{LUCJ}}}
\newcommand{\EvarT}{E_{\mathrm{var}}^{\mathrm{T}}}

\definecolor{RLDarkGreen}{HTML}{1d3c34}
\definecolor{RLMidGreen}{HTML}{006f62}
\definecolor{RLOrange}{HTML}{ff7500}

\definecolor{riverlane_green}{RGB}{0, 150, 143}
\definecolor{riverlane_orange}{RGB}{255, 117, 0}

\usepackage{hyperref}
\hypersetup{
  colorlinks   = true, %Colours links instead of ugly boxes
  urlcolor     = riverlane_green, %Colour for external hyperlinks
  linkcolor    = riverlane_green, %Colour of internal links
  citecolor    = riverlane_orange  %Colour of citations
}

\begin{document}

\title{A quantum computing approach to fixed-node Monte Carlo using classical shadows}

\author{Nick S. Blunt$^*$}
\affiliation{Riverlane, Cambridge, CB2 3BZ, UK}

\author{Laura Caune}
\affiliation{Riverlane, Cambridge, CB2 3BZ, UK}

\author{Javiera Quiroz-Fernandez}
\affiliation{Riverlane, Cambridge, CB2 3BZ, UK}
\affiliation{Department of Materials Science and Metallurgy, University of Cambridge, Cambridge CB3 0FS, United Kingdom}

\date{\today}

\begin{abstract}
Quantum Monte Carlo (QMC) methods are powerful approaches for solving electronic structure problems. Although they often provide high-accuracy solutions, the precision of most QMC methods is ultimately limited by a trial wave function that must be used. Recently, an approach has been demonstrated to allow the use of trial wave functions prepared on a quantum computer [Nature 603, 416 (2022)] in the auxiliary-field QMC (AFQMC) method, using classical shadows to estimate the required overlaps. However, this approach has an exponential post-processing step to construct these overlaps, when performing classical shadows obtained using random Clifford circuits. Here, we study an approach to avoid this exponentially scaling step by using a fixed-node Monte Carlo method, based on full configuration interaction quantum Monte Carlo (FCIQMC). This method is applied to the local unitary cluster Jastrow (LUCJ) ansatz. We consider H$_4$, ferrocene and benzene molecules using up to $12$ qubits as examples. Circuits are compiled to native gates for typical near-term architectures, and we assess the impact of circuit-level depolarizing noise on the method. We also provide a comparison of AFQMC and fixed-node approaches, demonstrating that AFQMC is more robust to errors, although extrapolations of the fixed-node energy reduce this discrepancy. Although the method can be used to reach chemical accuracy, the sampling cost to achieve this is high even for small active spaces, suggesting caution for the prospect of outperforming conventional QMC approaches.
\end{abstract}

\maketitle

\def\thefootnote{$^*$}\footnotetext{nick.blunt@riverlane.com}

\section{Introduction}
\label{sec:intro}

Quantum Monte Carlo (QMC) methods are among the most powerful methods for performing electronic structure calculations, often allowing high-accuracy estimates to be obtained for challenging systems where other methods are less reliable, and are widely used to provide benchmark results \cite{becca_sorella_2017}. Among these QMC methods, so-called projector Monte Carlo methods, which often attempt to perform imaginary-time propagation, are widely used. Exact imaginary-time evolution would allow the ground state to be sampled without error, but in practice QMC methods suffer from a sign problem which prevents such exact propagation. To avoid the sign problem, the evolution must be approximately guided, usually by means of a trial wave function, which is a ``best guess'' at the true solution. The accuracy of this trial wave function ultimately determines the accuracy of the final QMC estimates. Therefore, methods to optimize and make use of better trial wave functions are extremely valuable, and an important area of research.

Projector Monte Carlo methods perform sampling of some underlying states, $\{ | w_i \rangle \}$, such as Slater determinants. In order to use a given trial wave function $| \psiT \rangle$, it is necessary to calculate the corresponding overlaps, $\langle w_i | \psiT \rangle$, for each state $| w_i \rangle$ sampled during the algorithm. This provides a fundamental limitation on which trial wave functions can be used in practice; if $\langle w_i | \psiT \rangle$ is exponentially expensive to calculate then the use of $| \psiT \rangle$ is generally impractical. A further limitation is the ability to optimize the desired wave function parameterization, which for many wave function ansatze may be extremely challenging.

Recently, Huggins \emph{et al.} \cite{huggins_2022} demonstrated a novel approach to tackling these problems using quantum computation. In their approach, a trial wave function is prepared on a quantum computer, and the required overlaps are estimated using the classical shadows method \cite{huang_2020}. Given the very large number of overlaps that are typically required in a QMC simulation, the classical shadows approach is appealing because the number of samples required is only logarithmic in the number of overlaps to be estimated, in order to achieve a fixed precision in each overlap estimate with high probability. Moreover, all required overlaps can be estimated as a post-processing step, avoiding the need for constant communication between the classical and quantum processors. Their approach was applied for an example up to $16$ qubits, successfully obtaining energies within chemical accuracy. Despite the successful application of the method, a fundamental limitation was identified, namely that constructing each $\langle w_i | \psiT \rangle$ from the classical shadows has an exponential cost, when obtaining classical shadows over random Clifford circuits. This is due to the particular flavor of QMC used, auxiliary-field QMC (AFQMC) \cite{zhang_2003, motta_2018}, which puts a particularly strict criterion on the overlaps required.

It has since been shown that this exponential post-processing step can be avoided by using random Matchgate shadows instead of Clifford shadows \cite{wan_2023}. Recently, this approach was tested experimentally on a trapped ion quantum processor \cite{huang_2024}, where it was demonstrated that the resulting method has considerable tolerance to noise. However, the authors argue that the post-processing remains a formidable overhead, requiring thousands of CPU hours even for small chemical systems, and a high-polynomial-scaling to larger systems (although this scaling can likely be reduced by a number of techniques \cite{jiang_2024}). Ref.~\cite{kiser_2024} also studied the overlap cost of this scheme, coming to similar conclusions regarding scaling. So, while this approach remains promising, further development work is needed to reduce its cost, and it is valuable to investigate other approaches in parallel.

A separate approach to avoid this exponential-scaling step, as pointed out by the authors in \cite{huggins_2022}, is by using alternative QMC methods, particularly those based on the fixed-node approximation. In short, these QMC methods sample Slater determinants with a fixed orbitals basis and, as a consequence, the overlap estimation task is simplified and the necessary post-processing terms can be calculated in polynomial-scaling time. Previous studies have investigated the QC-QMC methodology with the fixed-node approximation \cite{xu_2023, mazzola_2022}. For example, Ref.~\cite{xu_2023} performed Green's functions Monte Carlo (GFMC), and proposed and tested a number of extensions, including the use of Bayesian inference to reduce the statistical error in estimates of $\langle w_i | \psiT \rangle$. However, such studies have focused on estimating the overlaps using Hadamard tests and related circuits, rather than the classical shadows-based approach proposed in \cite{huggins_2022}.

In this paper, we implement and investigate the QC-QMC methodology in combination with the fixed-node full configuration interaction quantum Monte Carlo (FCIQMC) method \cite{blunt_2021}. We implement this method using classical shadows to construct overlaps, sampling from ensembles of both $n$-qubit Cliffords circuits, and tensor products of single-qubit Clifford circuits. We test this approach for H$_4$, ferrocene and benzene molecules, in active spaces up to $6$ spatial orbitals, corresponding to quantum circuits with $12$ qubits. We make use of a recently-developed quantum ansatz, the local unitary cluster Jastrow (LUCJ) ansatz \cite{motta_2023, motta_2024}, which aims to bridge the gap between physically motivated wave functions, and those that can be reliably prepared on near-term quantum hardware. In this way, our work also provides further benchmarking for these quantum trial states. We compile the corresponding circuits to typical native gates for superconducting quantum processors, and test the methodology under circuit-level depolarizing errors. We also consider extensions to reduce the fixed-node error, including the use of extrapolations of the fixed-node parameter \cite{beccaria_2001}. Lastly, we conclude by comparing results from the fixed-node approximation to the phaseless AFQMC method.

We show that the fixed-node QC-QMC approach can significantly improve the variational energy estimate for a given trial wave function, in some cases removing over $90\%$ of error compared to VQE energies, and can obtain energies within a few mHa of exact result in the presence of both sampling noise and depolarizing errors. While this is promising, the sampling cost to achieve this is formidable, requiring more than $10^5$ Clifford shadow circuits to be performed for convergence, even in small chemical active spaces. As such, we expect that the methodology would require significant developments to be competitive with state-of-the-art QMC approaches.

The structure of the paper is as follows. In Section~\ref{sec:fn_fciqmc} we briefly introduce FCIQMC and the fixed-node approximation, before giving an overview of the classical shadows approach and its use in estimating overlaps in Section~\ref{sec:qc_qmc}. We then introduce the LUCJ ansatz in Sections~\ref{sec:lucj} and \ref{sec:real_trial_wf}, including its circuit compilation in QC-QMC. In the results, Section~\ref{sec:results}, we first show an application to H$_4$ in a minimal basis set, before investigating the effect of depolarizing errors. We then study two larger chemical active spaces, and finally conclude by comparing to AFQMC results for the H$_4$ example, including comparison with extrapolations of the fixed-node energy.

\section{Theory}
\label{sec:theory}

\subsection{Fixed-node FCIQMC}
\label{sec:fn_fciqmc}

The FCIQMC method \cite{booth_2009} is a type of projector QMC algorithm. The approach performs approximate imaginary-time propagation by applying the operator
\begin{equation}
    P = \mathbb{1} - \Delta \tau H
    \label{eq:propagator}
\end{equation}
to some initial state, where $H$ is the Hamiltonian, and $\Delta \tau$ is a time step for the propagation. Provided $1 / \Delta \tau$ is large enough compared to the spectral width of $H$, and $|\psi\rangle$ has overlap with the ground state of $H$, then $P^n |\psi\rangle$ converges to the exact ground state of $H$ in the limit of large $n$.

Applying $P$ exactly would require storing a vector in the full Hilbert space. Instead, FCIQMC performs a stochastic sampling of this propagation in an attempt to avoid this bottleneck. We do not present the algorithm to achieve this here, but refer to one of several explanations in the literature \cite{booth_2009, spencer_2012, petruzielo_2012, blunt_2015}. Like all projector QMC methods, FCIQMC suffers from a sign problem when performing ``free propagation'' in this manner \cite{spencer_2012}, resulting in an exponential cost to sample the exact ground-state wave function in general.

To avoid the sign problem, an approximation must be applied. In the context of FCIQMC, the most commonly applied approximation has been the initiator method \cite{cleland_2010, cleland_2011}, which does not make use of a trial wave function. In this paper we instead consider the fixed-node approximation with FCIQMC. This was first considered in \cite{kolodrubetz_2012} for lattice models, and extended to \emph{ab initio} chemical systems in \cite{blunt_2021}. The fixed-node approximation within FCIQMC is identical to the approximation more commonly applied in Green's function Monte Carlo \cite{becca_sorella_2017}, and the two resulting methods are closely related. However, in the context of FCIQMC, there is the additional option to partially lift the fixed-node approximation, which is possible due to the annihilation step that is performed in the FCIQMC method.

In the fixed-node method, the exact Hamiltonian is replaced by an approximate one that depends on a trial wave function. The fixed-node Hamiltonian is designed to ensure that the determinants sampled in a QMC simulation (called ``walkers'') always have the same sign as the trial wave function. Because walkers on the same determinant always have the same sign, the sign problem is removed, and the ground state of the fixed-node Hamiltonian can be sampled efficiently.

We denote the trial wave function $| \psiT \rangle$, and a complete set of basis states by $\{ | D_i \rangle \}$. In this work, these basis states will always be Slater determinants, but other states can be used in general. The trial wave function can be expressed in the determinant basis as
\begin{equation}
    | \psiT \rangle = \sum_i \psiiT | D_i \rangle,
\end{equation}
where $\psiiT = \langle D_i | \psiT \rangle$ denotes the overlap of the trial wave function with a given Slater determinant, which will be a key quantity in this study. We also define
\begin{equation}
    s_{ij} = \psiiT H_{ij} \psijT,
    \label{eq:sij}
\end{equation}
where $H_{ij} = \langle D_i | H | D_j \rangle$. The importance of $s_{ij}$ is that it can be used to check if a sign violation is introduced by $H$, relative to $|\psiT\rangle$, when spawning a walker from $|D_j\rangle$ to $|D_i\rangle$. As in Eq.~\ref{eq:propagator}, the Hamiltonian appears with a minus sign in $P$, and so if $s_{ij} > 0$ then a sign violation can occur between determinants $|D_i\rangle$ and $|D_j\rangle$ when applying $P$. This must be addressed to avoid a sign problem, allowing stable propagation without requiring an exponentially large walker population.

The fixed-node Hamiltonian is defined by \cite{van_bremmel_1994, becca_sorella_2017, sorella_2000}
\begin{equation}
   \Hfn(\gamma) = \sum_{ij} H_{ij}^{\mathrm{fn}}(\gamma) | D_i \rangle \langle D_j |,
\end{equation}
with
\begin{equation}
H_{ij}^{\mathrm{fn}}(\gamma) = \begin{cases}
  H_{ii} + (1 + \gamma)\mathcal{V}_i^{\mathrm{sf}}, & \text{if $i = j$},\\
  H_{ij}, & \text{if $i \ne j$, $s_{ij} < 0$}, \\
  -\gamma H_{ij}, & \text{if $i \ne j$, $s_{ij} > 0$},
  \end{cases}
  \label{eq:fn_hamiltonian}
\end{equation}
and $\gamma \in \mathbb{R}$. This is the form of the fixed-node approximation introduced by van Bremmel \emph{et al.} \cite{van_bremmel_1994}. The quantity
\begin{equation}
    \mathcal{V}_i^{\mathrm{sf}} = \sum_{j: s_{ij} > 0} H_{ij} \frac{\psijT}{\psiiT}
\end{equation}
is known as the sign-flip potential. Note that $\Hfn(-1) = H$, so that the original Hamiltonian can be obtained with $\gamma=-1$. The Hamiltonian is sign-problem-free for all $\gamma \ge 0$, but the fixed-node approximation is most commonly concerned with $\gamma = 0$ in particular. We refer to the ground-state energy of $\Hfn(\gamma)$ as the fixed-node energy, denoted $\Efn(\gamma)$. For ease of reference we use this notation generally, though technically $\Hfn(\gamma)$ is only ``fixed-node'' for $\gamma \ge 0$.

Despite the simplicity of this approach to removing sign violations, the fixed-node energy is variational (for $\gamma \ge -1$), and also guaranteed to be lower than the variational energy of the trial wave function,
\begin{equation}
    E_{\mathrm{var}} = \frac{\langle \psiT | H | \psiT \rangle }{ \langle \psiT | \psiT \rangle}.
    \label{eq:var_estimator}
\end{equation}

In fixed-node FCIQMC, we perform FCIQMC by applying the operator
\begin{equation}
    \Pfn = \mathbb{1} - \Delta \tau \Hfn.
    \label{eq:propagator_fn}
\end{equation}
If $| \Psi_{\textrm{QMC}} \rangle = \sum_i C_i | D_i \rangle$ is the FCIQMC wave function at a given iteration, where $\{ C_i \}$ are the amplitudes of the walkers, the ground-state energy of $\Hfn$ may be estimated by
\begin{equation}
    E = \frac{\sum_i C_i \EiL}{\sum_i C_i},
    \label{eq:fn_energy_estimator}
\end{equation}
where
\begin{equation}
    \EiL = \sum_j H_{ij} \frac{\psijT}{\psiiT}
    \label{eq:local_energy}
\end{equation}
is referred to as the ``local energy'' on determinant $|D_i\rangle$. Eq.~\ref{eq:fn_energy_estimator} is an energy estimate from a single FCIQMC iteration; the final estimate is obtained by averaging over a large number of such iterations. Note that the correctness of Eq.~\ref{eq:fn_energy_estimator} depends on also applying importance sampling, which is described in \cite{blunt_2021}.

Lastly we note that it was proven in \cite{beccaria_2001} that $\Efn(\gamma)$ is a concave function in $\gamma$. Therefore, performing a linear extrapolation to $\gamma=-1$ using any two estimates $\Efn(\gamma_1)$ and $\Efn(\gamma_2)$, for $\gamma_1 > -1$ and $\gamma_2 > -1$, will result in both an improved and variational estimate of the true ground-state energy. We will present results in this paper to demonstrate the improvement from such extrapolations in practice.

For further details of the fixed-node QMC approach used in this paper, including importance sampling, we refer to \cite{blunt_2021}.

\subsection{Classical shadows and the QC-QMC method}
\label{sec:qc_qmc}

The key object that determines the accuracy of the fixed-node approach is the trial wave function, $| \psiT \rangle$. From Section \ref{sec:fn_fciqmc}, it can be seen that this trial state enters the QMC simulation in two places; first in the definition of the fixed-node Hamiltonian, Eq. \ref{eq:fn_hamiltonian} and second in the local energy calculation, Eq. \ref{eq:local_energy}. Crucially, in both of these cases, the trial wave function enters through its overlaps, $\psiiT$, with determinants. Therefore, being able to calculate $\psiiT$ efficiently is a key task. In traditional fixed-node Monte Carlo approaches, the form of $| \psiT \rangle$ is limited by this requirement. For some ansatze, calculating $\langle D_i | \psiT \rangle$ has an exponential cost, and the use of such ansatze is generally avoided.

Huggins \emph{et al}. suggested quantum computation as an approach to use a larger class of trial wave functions, and potentially improve the accuracy of projector QMC methods, including AFQMC and fixed-node Monte Carlo \cite{huggins_2022}. There is now a large literature on how to prepare various quantum states as quantum circuits, that are challenging to prepare classically \cite{peruzzo_2014, kandala_2017, motta_2023, tilly_20221}. For a prepared quantum state, estimating $\langle D_i | \psiT \rangle$ can be achieved with a Hadamard test. However, if the number of overlaps to estimate is denoted $M$, then $\mathcal{O}(M)$ Hadamard tests will be required, and communication between the CPU and QPU will be required for each overlap. Instead, Huggins \emph{et al}. suggest using the classical shadows approach, where it is possible to estimate $M$ observables to precision $\epsilon$ with high probability by using $\mathcal{O}(\mathrm{log}(M)/\epsilon^2)$ samples. Moreover, all of the required measurements on the QPU can be obtained prior to the classical QMC simulation, avoiding the need for communication between the CPU and QPU. This approach is referred to as the quantum-classical hybrid QMC (QC-QMC) method.

We now briefly review the classical shadows approach for estimating a general observable, as introduced in \cite{huang_2020}, and then its specialization to estimating overlaps.

Consider the task of calculating observables for the trial state $|\psiT\rangle$, which can be prepared by a quantum circuit. We denote the corresponding density operator $\rho = | \psiT \rangle \langle \psiT |$. The classical shadows protocol is concerned with estimating expectation values $o_j = \mathrm{Tr}[\rho O_j]$ for observables $O_j$. This is achieved by preparing $ | \psiT \rangle$ and applying random unitaries, $U_k$, sampled uniformly from an ensemble $\mathcal{U}$, and measuring in the computational basis to obtain a bitstring $|b_k\rangle$. This process is repeated for a large number of random unitaries, and each $U_k$ and $|b_k\rangle$ is stored classically. Consider the expectation value of $U_k^{\dagger} |b_k \rangle \langle b_k | U_k$ over the samples taken, which can be viewed as a quantum channel acting on $\rho$, denoted $\mathcal{M}(\rho)$,
\begin{equation}
    \mathcal{M}(\rho) = \ex_k \Big[ \, U_k^{\dagger} |b_k \rangle \langle b_k | U_k \, \Big].
\end{equation}
For appropriate ensembles $\mathcal{U}$, there exists a unique inverse, $\mathcal{M}^{-1}(\cdot)$, with a simple expression, allowing one to write
\begin{equation}
    \rho = \ex_k \Big[ \, \mathcal{M}^{-1} \Big( U_k^{\dagger} |b_k \rangle \langle b_k | U_k \Big) \, \Big].
    \label{eq:rho_from_shadows}
\end{equation}
A single estimate $\mathcal{M}^{-1} ( \, U_k^{\dagger} |b_k \rangle \langle b_k | U_k \, )$ is viewed as a ``classical snapshot'' of the state $\rho$, denoted $\rho_k$. Expectation values of the form  $\mathrm{Tr}[\rho O_j]$ can be estimated by averaging $\mathrm{Tr}[\rho_k O_j]$ over all snapshots. This provides an unbiased estimator for the corresponding observable.

Specifically, Huang \emph{et al}. consider two ensembles, $\mathcal{U}$; (i) the group of $n$-qubit Clifford unitaries, and (ii) the group of tensor products of $1$-qubit Clifford unitaries. The corresponding quantum channels are labeled $\mathcal{M}_n$ and $\mathcal{M}_P$, respectively. Ref~\cite{huang_2020} proves that the corresponding inverses are
\begin{align}
    \mathcal{M}_n^{-1}(X) &= (2^n + 1)X - \mathbb{1}, \label{eq:m_n_inverse} \\
    \mathcal{M}_P^{-1}(X) &= \bigotimes_{i=1}^n \mathcal{M}_1^{-1}(X).
\end{align}
This allows simple expressions for unbiased estimators of observables, in terms of the sampled $U_k$ and $|b_k\rangle$.

Next we consider the specific case of estimating overlaps of the form $\langle D_i | \psiT \rangle$, as required for QC-QMC. To write such an overlap in the form of an expectation value amenable to the classical shadows approach, we consider preparing the following state
\begin{equation}
    | \tau \rangle = \frac{1}{\sqrt{2}} ( | 0 \rangle + | \psiT \rangle),
\end{equation}
where $| 0 \rangle$ is the computational basis state of all $0$'s. More generally, $|0\rangle$ can be any state which is orthogonal to $| \psiT \rangle$. Since $| \psiT \rangle$ will be prepared to represent an $N$-particle quantum state, under the Jordan-Wigner transformation it will be mapped to a linear combination of computational basis states with Hamming weight $N$ (assuming that the ansatz does not break particle number symmetry), and will be orthogonal to $|0\rangle$.

We now let $\rho$ represent the density operator of the state $| \tau \rangle$, i.e. $\rho = | \tau \rangle \langle \tau|$. Under the above orthogonality condition,
\begin{equation}
    \langle D_i | \psiT \rangle =  2 \, \textrm{Tr} \big[ \, | 0 \rangle \langle D_i | \, \rho \, \big].
    \label{eq:overlap_eq_1}
\end{equation}
This puts the overlap in a form where the classical shadows protocol can be applied. We emphasize that the quantum circuit employed in QC-QMC must prepare the state $( | 0 \rangle + | \psiT \rangle) / \sqrt{2}$, not the state $| \psiT \rangle$. For many trial states, including the ones used in this paper, this can be achieved with minimal additional depth compared to the circuit required to prepare $| \psiT \rangle$, though this may not always be true.

Combining Eqs.~\ref{eq:rho_from_shadows} and \ref{eq:overlap_eq_1} gives
\begin{equation}
    \langle D_i | \psiT \rangle = 2 \,\ex_k \Big[ \textrm{Tr} \big[ \, |0 \rangle \langle D_i | \, \mathcal{M}^{-1} \Big( U_k^{\dagger} |b_k \rangle \langle b_k | U_k \Big) \, \big] \Big].
\end{equation}

The final expression used to estimate overlaps for the QMC method is then obtained by using the appropriate $\mathcal{M}^{-1}(\cdot)$, which will depend on the ensemble $\mathcal{U}$ used. For example, considering the ensemble of $n$-qubit Cliffords, and therefore taking $\mathcal{M}_n^{-1}(\cdot)$ as in Eq.~\ref{eq:m_n_inverse},
\begin{equation}
        \langle D_i | \psiT \rangle = 2(2^n + 1) \ex_k \Big[ \langle D_i | U_k^{\dagger} |b_k \rangle \langle b_k | U_k | 0 \rangle \Big].
        \label{eq:overlap_estimator}
\end{equation}
In the AFQMC method, the walkers $| D_i \rangle$ can be general Slater determinants in an arbitrary orbital basis. These do not map to a polynomial-sized linear combination of stabilizer states, and so the term $ \langle D_i | U_k^{\dagger} |b_k \rangle$ cannot be calculated in polynomial time. In contrast, the FCIQMC and GFMC methods can work in a basis of determinants with fixed orbitals, and so each $| D_i \rangle$ is a computational basis state, and $ \langle D_i | U_k^{\dagger} |b_k \rangle$ can be calculated in $\mathcal{O}(n^3)$ time \cite{aaronson_2004}. Each expectation value can then be estimated by averaging over all classical snapshots, $\{U_k, |b_k\rangle \}$. The most computationally challenging task in this process is the preparation of the trial state on the quantum computer, which determines the distribution of the $\{ |b_k\rangle \}$ sampled, but the remaining steps are classically ``easy'' to perform.

\subsection{The local unitary cluster Jastrow ansatz}
\label{sec:lucj}

In this study we perform QC-QMC with the local unitary cluster Jastrow (LUCJ) ansatz \cite{motta_2023, motta_2024}, which we briefly review. Although we will only perform simulated results in this study, we choose to investigate an example that can be compiled to realistic near-term hardware and a square qubit lattice.

First consider the following ansatz, known as the unitary cluster Jastrow (UCJ) ansatz \cite{matsuzawa_2020},
\begin{equation}
    | \Psi_{\mathrm{UCJ}} \rangle = \prod_{\mu=1}^L e^{K_{\mu}} e^{iJ_{\mu}} e^{-K_{\mu}} |D_0\rangle.
\end{equation}
This ansatz can be motivated as a unitarization of cluster-Jastrow wave functions \cite{neuscamman_2013, neuscamman_2013_2}, which have been used in variational Monte Carlo (VMC) simulations. Here, each $e^{-K_{\mu}}$ is an orbital rotation operator, each $e^{iJ_{\mu}}$ is a unitary Jastrow factor, and $| D_0 \rangle$ is a reference Slater determinant. In general this ansatz consists of $L$ layers of rotations and Jastrow factors, so that the ansatz can be systematically improved by increasing $L$. In this work, we will focus on the $L=1$ case, and so drop the $\mu$ subscript from now on. Then we have,
\begin{equation}
    K = \sum_{pq,\sigma} K_{pq} \hat{a}_{p \sigma}^{\dagger} \hat{a}_{q \sigma}^{\dagger},
\end{equation}
\begin{equation}
    J = \sum_{pq, \sigma \tau} J_{pq, \sigma \tau} \hat{n}_{p \sigma} \hat{n}_{q \tau},
    \label{eq:general_jastrow}
\end{equation}
where $\hat{a}_{p\sigma}^{\dagger}$ is the creation operator for spatial orbital $p$ with spin label $\sigma$, and $\hat{n}_{p\sigma} = \hat{a}_{p\sigma}^{\dagger} \hat{a}_{p\sigma}$ is the corresponding number operator. $J_{pq,\sigma\tau}$ is real and Hermitian, while $K_{pq}$ is complex and anti-Hermitian.

\begin{figure*}[t]

\begin{tikzpicture}

\tikzset{dot/.style={fill=black,circle,inner sep=1.5pt}}

\draw[step=1.0cm,black, thick] (0,0) grid (3,1);

\draw (-1.5,0.5) node{a)};

\draw (0.0,1.4) node{$0$};
\draw (1.0,1.4) node{$1$};
\draw (2.0,1.4) node{$2$};
\draw (3.0,1.4) node{$3$};

\draw (0.0,-0.5) node{$4$};
\draw (1.0,-0.5) node{$5$};
\draw (2.0,-0.5) node{$6$};
\draw (3.0,-0.5) node{$7$};

\node[dot] at (0,0){};
\node[dot] at (1,0){};
\node[dot] at (2,0){};
\node[dot] at (3,0){};
\node[dot] at (0,1){};
\node[dot] at (1,1){};
\node[dot] at (2,1){};
\node[dot] at (3,1){};

\end{tikzpicture}
\hspace{1.5cm}
\begin{tikzpicture}
\tikzset{dot/.style={fill=black,circle,inner sep=1.5pt}}

\draw[step=1.0cm,black, thick] (0,0) grid (3,1);

\draw (0.0,1.4) node{$(0,\uparrow)$};
\draw (1.0,1.4) node{$(1,\uparrow)$};
\draw (2.0,1.4) node{$(2,\uparrow)$};
\draw (3.0,1.4) node{$(3,\uparrow)$};

\draw (0.0,-0.5) node{$(0,\downarrow)$};
\draw (1.0,-0.5) node{$(1,\downarrow)$};
\draw (2.0,-0.5) node{$(2,\downarrow)$};
\draw (3.0,-0.5) node{$(3,\downarrow)$};

\node[dot] at (0,0){};
\node[dot] at (1,0){};
\node[dot] at (2,0){};
\node[dot] at (3,0){};
\node[dot] at (0,1){};
\node[dot] at (1,1){};
\node[dot] at (2,1){};
\node[dot] at (3,1){};

\end{tikzpicture}

\vspace{4mm}

b)
\begin{quantikz}[column sep={1mm}, row sep={2mm}]
 0 \hspace{2mm} & \qw &[5mm] \gate[2]{G}\gategroup[8,steps=5,style={dashed,rounded corners,fill=RLMidGreen!20, inner xsep=2pt},background]{} &[3mm] \qw         &[3mm] \gate[2]{G} &[3mm] \qw         &[3mm] \gate{\varphi} &[6mm] \qw & \ctrl{4}\gategroup[8,steps=7,style={dashed,rounded corners,fill=RLOrange!20, inner xsep=2pt},background]{} & \qw      & \qw      & \qw      &[3mm] \gate{\varphi} &[2mm] \ctrl{1} & \qw      &[6mm] \gate{\varphi}\gategroup[8,steps=5,style={dashed,rounded corners,fill=RLMidGreen!20, inner xsep=2pt},background]{} \qw &[3mm] \qw         &[3mm] \gate[2]{G} &[3mm] \qw         &[3mm] \gate[2]{G} &[5mm] \qw \\
 1 \hspace{2mm} & \qw &[5mm]             &[3mm] \gate[2]{G} &[3mm]             &[3mm] \gate[2]{G} &[3mm] \gate{\varphi} &[6mm] \qw & \qw      & \ctrl{4} & \qw      & \qw      &[3mm] \gate{\varphi} &[2mm]  \gate{\varphi} & \ctrl{1} &[6mm] \gate{\varphi} \qw &[3mm] \gate[2]{G} &[3mm]             &[3mm] \gate[2]{G} &[3mm]             &[5mm] \qw \\
 2 \hspace{2mm} & \qw &[5mm] \gate[2]{G} &[3mm]             &[3mm] \gate[2]{G} &[3mm]             &[3mm] \gate{\varphi} &[6mm] \qw & \qw      & \qw      & \ctrl{4} & \qw      &[3mm] \gate{\varphi} &[2mm]  \ctrl{1} & \gate{\varphi} &[6mm] \gate{\varphi} \qw &[3mm]             &[3mm] \gate[2]{G} &[3mm]             &[3mm] \gate[2]{G} &[5mm] \qw \\
 3 \hspace{2mm} & \qw &[5mm]             &[3mm] \qw         &[3mm]             &[3mm] \qw         &[3mm] \gate{\varphi} &[6mm] \qw & \qw      & \qw      & \qw      & \ctrl{4} &[3mm] \gate{\varphi} &[2mm]  \gate{\varphi} & \qw      &[6mm] \gate{\varphi} \qw &[3mm] \qw         &[3mm]             &[3mm] \qw         &[3mm]             &[5mm] \qw \\
 4 \hspace{2mm} & \qw &[5mm] \gate[2]{G} &[3mm] \qw         &[3mm] \gate[2]{G} &[3mm] \qw         &[3mm] \gate{\varphi} &[6mm] \qw & \gate{\varphi} & \qw      & \qw      & \qw      &[3mm] \gate{\varphi} &[2mm]  \ctrl{1} & \qw      &[6mm] \gate{\varphi} \qw &[3mm] \qw         &[3mm] \gate[2]{G} &[3mm] \qw         &[3mm] \gate[2]{G} &[5mm] \qw \\
 5 \hspace{2mm} & \qw &[5mm]             &[3mm] \gate[2]{G} &[3mm]             &[3mm] \gate[2]{G} &[3mm] \gate{\varphi} &[6mm] \qw & \qw      & \gate{\varphi} & \qw      & \qw      &[3mm] \gate{\varphi} &[2mm]  \gate{\varphi} & \ctrl{1} &[6mm] \gate{\varphi} \qw &[3mm] \gate[2]{G} &[3mm]             &[3mm] \gate[2]{G} &[3mm]             &[5mm] \qw \\
 6 \hspace{2mm} & \qw &[5mm] \gate[2]{G} &[3mm]             &[3mm] \gate[2]{G} &[3mm]             &[3mm] \gate{\varphi} &[6mm] \qw & \qw      & \qw      & \gate{\varphi} & \qw      &[3mm] \gate{\varphi} &[2mm]  \ctrl{1} & \gate{\varphi} &[6mm] \gate{\varphi} \qw &[3mm]             &[3mm] \gate[2]{G} &[3mm]             &[3mm] \gate[2]{G} &[5mm] \qw \\
 7 \hspace{2mm} & \qw &[5mm]             &[3mm] \qw         &[3mm]             &[3mm] \qw         &[3mm] \gate{\varphi} &[6mm] \qw & \qw      & \qw      & \qw      & \gate{\varphi} &[3mm] \gate{\varphi} &[2mm]  \gate{\varphi} & \qw      &[6mm] \gate{\varphi} \qw &[3mm] \qw         &[3mm]             &[3mm] \qw         &[3mm]             &[5mm] \qw
\end{quantikz}

\vspace{7mm}

c)
\begin{quantikz}[column sep={3mm}, row sep={1mm}]
\lstick{$\ket{0}$} & \gate{H}\gategroup[8,steps=3,style={dashed,rounded corners, inner xsep=2pt},background, label style={label position=below,anchor=north,yshift=-0.6cm}]{Prepare $\frac{1}{\sqrt{2}}(|0\rangle + |D_0\rangle)$} & \ctrl{4} & \ctrl{1} &[5mm] \gate[8, style={fill=RLMidGreen!20}][2cm]{e^{-K}}\gategroup[8,steps=3,style={dashed,rounded corners, inner xsep=2pt},background, label style={label position=below,anchor=north,yshift=-0.6cm}]{Apply orbital rotations and Jastrow} & \gate[8, style={fill=RLOrange!20}][2cm]{e^{iJ}} & \gate[8, style={fill=RLMidGreen!20}][2cm]{e^{K}} &[5mm] \gate[8][2cm]{\mathcal{U}}\gategroup[8,steps=2,style={dashed,rounded corners, inner xsep=2pt},background, label style={label position=below,anchor=north,yshift=-0.6cm}]{Random Clifford $\&$ measurement} & \meter{} \\
\lstick{$\ket{0}$} & \qw      & \qw      & \targ{}  &[5mm]                 &                  &                  &[3mm]          & \meter{} \\
\lstick{$\ket{0}$} & \qw      & \qw      & \qw      &[5mm]                 &                  &                  &[3mm]          & \meter{} \\
\lstick{$\ket{0}$} & \qw      & \qw      & \qw      &[5mm]                 &                  &                  &[3mm]          & \meter{} \\
\lstick{$\ket{0}$} & \qw      & \targ{}  & \ctrl{1} &[5mm]                 &                  &                  &[3mm]          & \meter{} \\
\lstick{$\ket{0}$} & \qw      & \qw      & \targ{}  &[5mm]                 &                  &                  &[3mm]          & \meter{} \\
\lstick{$\ket{0}$} & \qw      & \qw      & \qw      &[5mm]                 &                  &                  &[3mm]          & \meter{} \\
\lstick{$\ket{0}$} & \qw      & \qw      & \qw      &[5mm]                 &                  &                  &[3mm]          & \meter{}
\end{quantikz}
\caption{Circuit for the LUCJ ansatz for a H$_4$ example with $8$ qubits. (a) shows the assumed square-lattice qubit connectivity, including the qubit indices (left) and corresponding spin orbital labels (right). The structure of the LUCJ ansatz is shown in (b), with orbital rotation operators in green, and the unitary Jastrow factor in orange. The full circuit structure is shown in (c), including initial gates to prepare the GHZ state, and a final unitary sampled uniformly from $\mathcal{U}$, and subsequent measurement.}
\label{fig:lattice_and_circuits}
\end{figure*}
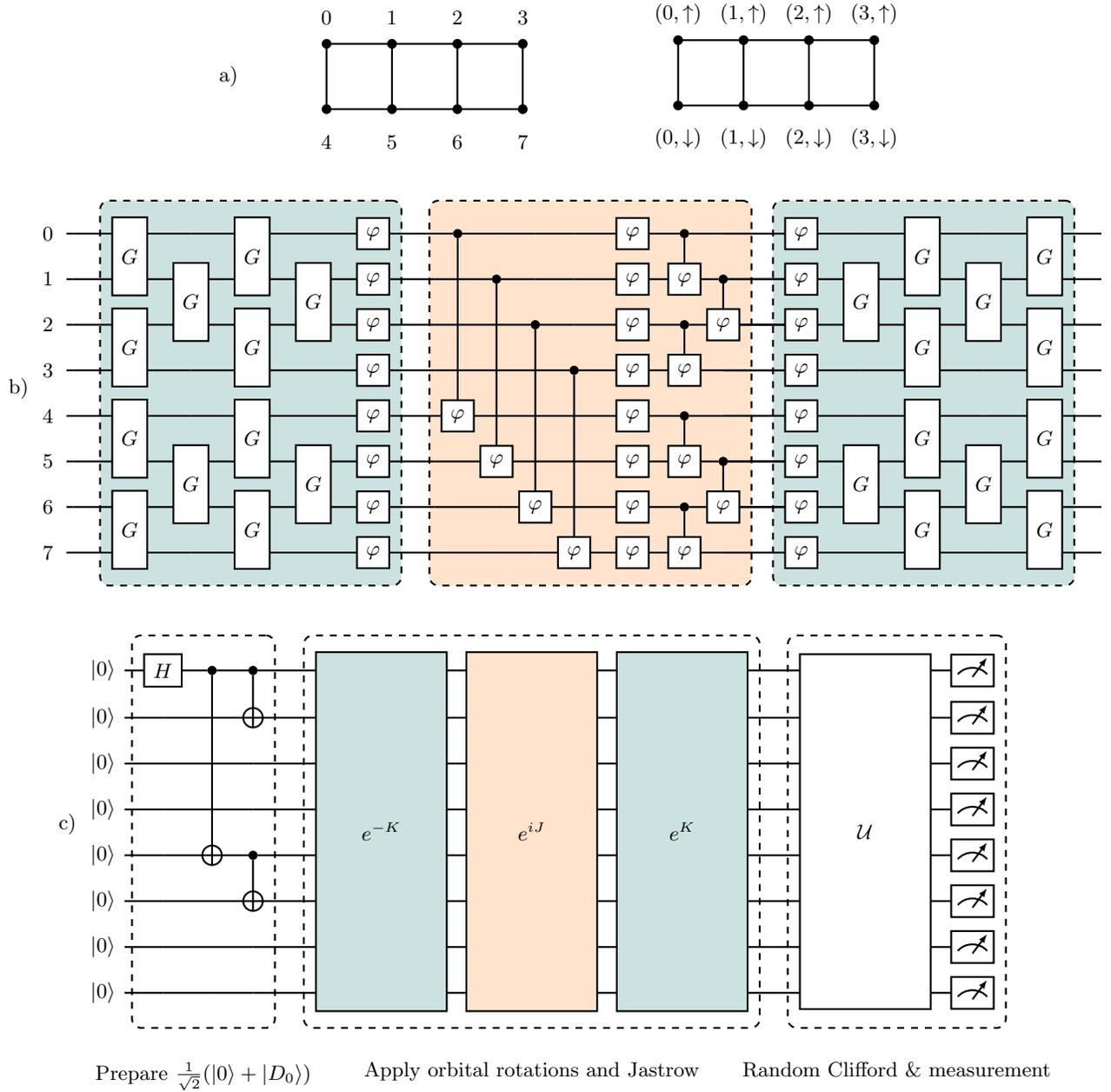

When mapping this ansatz to a quantum circuit representation, we will assume the Jordan-Wigner mapping throughout, so that a wave function for a system of $M$ spin orbitals is represented by $M$ qubits.

To assess how such an ansatz might be implemented on a near-term device, let us first consider the unitary Jastrow factor. Consider a single term $e^{iJ_{pq, \sigma \tau} \hat{n}_{p\sigma} \hat{n}_{q\tau}}$ for two different orbitals, $p\sigma \ne q\tau$. Under the Jordan-Wigner transformation we have
\begin{equation}
    e^{iJ_{kl} \hat{n}_k \hat{n}_l} = \mathrm{exp}\Big[\frac{i J_{kl}}{4} ( \mathbb{1} - Z_k - Z_l + Z_k Z_l )\Big].
\end{equation}
This corresponds to a standard CPHASE gate, which is a native gate for certain superconducting quantum processors, provided that the qubits are neighboring. However, the Jastrow factor in Eq.~\ref{eq:general_jastrow} couples together all pairs of qubits. Therefore, this either requires a device with all-to-all connectivity, or in general the use of swap network with $\mathcal{O}(N^2)$ SWAP gates.

To avoid this high cost for near-term devices, on can instead consider a Jastrow factor that only couples pairs of qubits connected on the device. This approach was used in \cite{huggins_2022}, where for H$_4$ the authors considered multiple hardware efficient layers with orbital rotations in-between. Ref.~\cite{motta_2023} formalized and developed this idea, which they call the \emph{local} unitary cluster Jastrow (LUCJ) ansatz. The hardware-efficient restriction inevitably reduces the accuracy of the Jastrow factor compared to the full UCJ ansatz, however it also reduces the cost of implementation. In this work we will consider the case of a square lattice, where the orbitals are mapped to qubits such that spin-up orbitals are a single row of the lattice, while the spin-down orbitals are mapped to a second row. This is shown in Fig.~\ref{fig:lattice_and_circuits}(a) for an example with 8 spin orbitals, showing the qubit labeling and corresponding spin-orbital labels from the Jordan-Wigner mapping.

The local unitary Jastrow factor can then be represented by a product of CPHASE gates between pairs of qubits corresponding to both opposite-spin and same-spin orbitals, in addition to a layer of PHASE or RZ gates to account for the diagonal $J_{pp,\uparrow\uparrow}$ and $J_{pp,\downarrow\downarrow}$ terms. An example is shown in Fig.~\ref{fig:lattice_and_circuits} for a system with $8$ spin orbitals.

The orbital rotation operator $e^{-K}$ can be represented by $M$ layers of Givens rotations, $M(M-1)$ in total, together with a single layer of rotation or PHASE gates \cite{jiang_2018, clements_2016}. An example is again shown in Fig.~\ref{fig:lattice_and_circuits}. A Givens rotation, acting on qubits in the Jordan-Wigner representation, can be defined as
\begin{equation}
G(\theta, \beta) = 
\begin{pmatrix}
1 & 0 & 0 & 0\\[0.2em]
0 & \mathrm{cos}(\frac{\theta}{2}) & -i \, \mathrm{sin}(\frac{\theta}{2}) \, e^{-i\beta} & 0\\[0.4em]
0 & -i \, \mathrm{sin}(\frac{\theta}{2}) \, e^{i\beta} & \mathrm{cos}(\frac{\theta}{2}) & 0\\[0.2em]
0 & 0 & 0 & 1
\end{pmatrix},
\end{equation}
which can be expressed in terms of a small number of one- and two-qubit gates on superconducting devices, for example.

\subsection{Fixed-node trial wave function from LUCJ ansatz}
\label{sec:real_trial_wf}

The overlaps $\langle D_i | \LUCJ \rangle$ for the LUCJ ansatz will be complex in general. However, for the fixed-node approximation, it is necessary to use a real trial wave function. Simply taking the real part of the wave function can either improve or worsen the wave function in general. Moreover, if overlaps happen to have a large imaginary component and small real component, then the classical shadows procedure will require more measurements to achieve a given relative precision in $\mathrm{Re}[\langle D_i | \LUCJ \rangle]$. To avoid this latter situation, we simply choose a reference determinant $|D_0\rangle$ for which the magnitude of $r e^{i\theta_0} = \langle D_0 | \LUCJ \rangle$ is known to be relatively large, and then apply a global phase $e^{-i\theta_0}$ to the wave function such that $\langle D_0 | \LUCJ \rangle e^{-i\theta_0}$ is real. Here, $\theta_0$ is estimated from $\langle D_0 | \LUCJ \rangle$ obtained by the classical shadows procedure. We then work with the real part of the subsequent wave function. We find this simple procedure to work well in practice, and that it often improves the variational energy of the original LUCJ wave function. For notational simplicity, we will refer to the latter wave function as $| \psiT \rangle$ throughout, and the original complex LUCJ wave function as $ | \LUCJ \rangle $, so that
\begin{equation}
    \langle D_i | \psiT \rangle = \mathrm{Re}[\langle D_i | \LUCJ \rangle e^{-i\theta_0}].
    \label{eq:real_component}
\end{equation}
From a  symmetries point of view, this procedure enforces time-reversal symmetry, which is broken in the LUCJ ansatz. Note however that this procedure will give higher energies than taking the real part of the wave function \emph{before} performing optimization, as the former is in the variational optimization space of the latter.

When considering the variational energy estimator of the form $\frac{ \langle \psi | H | \psi \rangle }{ \langle \psi | \psi \rangle }$, we will use the notation $\EvarLUCJ$ and $\EvarT$ to distinguish between the variational energies of the LUCJ and final trial wave functions.

\subsection{Random Clifford ensembles}
\label{sec:random_cliffords}

In our simulations, we will consider different distributions of random Cliffords for $\mathcal{U}$. In particular, for a circuit of $n$ qubits, we use: (i) tensor products of random single-qubit Cliffords, (ii) tensor products of random $n/2$-qubit Cliffords and (iii) random $n$-qubit Cliffords. For ease of reference, we will sometimes refer to the random ensembles as $\mathcal{C}_1^{\otimes n}$, $\mathcal{C}_{n/2}^{\otimes 2}$ and $\mathcal{C}_n$, respectively. However note that in practice we sample from a smaller but equivalent set; for example, circuits for ensemble (i) are implemented by equivalently measuring in a random Pauli basis. Following \cite{huggins_2022}, we will sometimes refer to shadows from ensemble (ii) as ``partitioned'' classical shadows. We give additional details on the sampling and compilation of circuits for ensembles (ii) and (iii) in Section~\ref{sec:noise}, where simulations are performed under circuit-level noise; in that case, compilation details will affect results, whereas noiseless simulations considered in Sections~\ref{sec:h4} and \ref{sec:ferrocene_benzene} will be independent of compilation details.

\section{Results}
\label{sec:results}

\subsection{Implementation details}
\label{sec:implementation}

Simulations are performed for three chemical systems, linear H$_4$, ferrocene and benzene, with $8$, $10$ and $12$ spin orbitals respectively. Linear H$_4$ calculations used a STO-6G basis and internuclear distance of $2$~$a_0$. For ferrocene, we used the atomic valence active space (AVAS) \cite{avas_paper} method to generate an active space of 6 electrons in 5 spatial orbitals. Following \cite{avas_paper}, in the AVAS method we first performed restricted open-shell Hartree--Fock (ROHF) in a cc-pVTZ-DK basis, including scalar relativistic effects by the exact-two-component (X2C) method. A threshold of $0.5$ was then used to select the active space via the AVAS approach. For benzene, we performed RHF in a 6-31G basis and then selected an active space of $6$ electrons in the $6$ valence $\pi$ and $\pi^*$ orbitals. For each system, the orbitals were then localized, before generating one- and two-body integrals for the subsequent calculations. All of the above steps were performed using PySCF \cite{pyscf, pyscf_2}. OpenFermion \cite{openfermion} was used to generate the qubit Hamiltonian via the Jordan-Wigner mapping. The LUCJ ansatz for each system was optimized by minimizing the variational energy using the JAX library \cite{jax_github} and BFGS algorithm. Fixed-node FCIQMC and AFQMC calculations were performed using the Dice code \cite{dice_github}. The implementation of FCIQMC from this code is described in \cite{blunt_2021}, and the implementation of AFQMC in \cite{mahajan_2021, mahajan_2022}.

Quantum circuits were simulated using pyQuil and associated quantum virtual machines (QVMs) \cite{smith_2016}. Before running these simulations, we compiled all circuits to a native gate set consisting of single qubit gates (RX and RZ) and two-qubit gates (CZ, CPHASE and XY). For each random Clifford generated, we perform just a single repetition, or ``shot'', of the circuit. In the results, the number of classical shadows that we average over to obtain $\langle D_i | \psiT \rangle$ estimates from Eq.~\ref{eq:overlap_estimator} is referred to as the ``number of shadow circuits'' performed.

\subsection{Demonstration with H$_4$}
\label{sec:h4}

\begin{figure}[t]
\includegraphics[width=\linewidth]{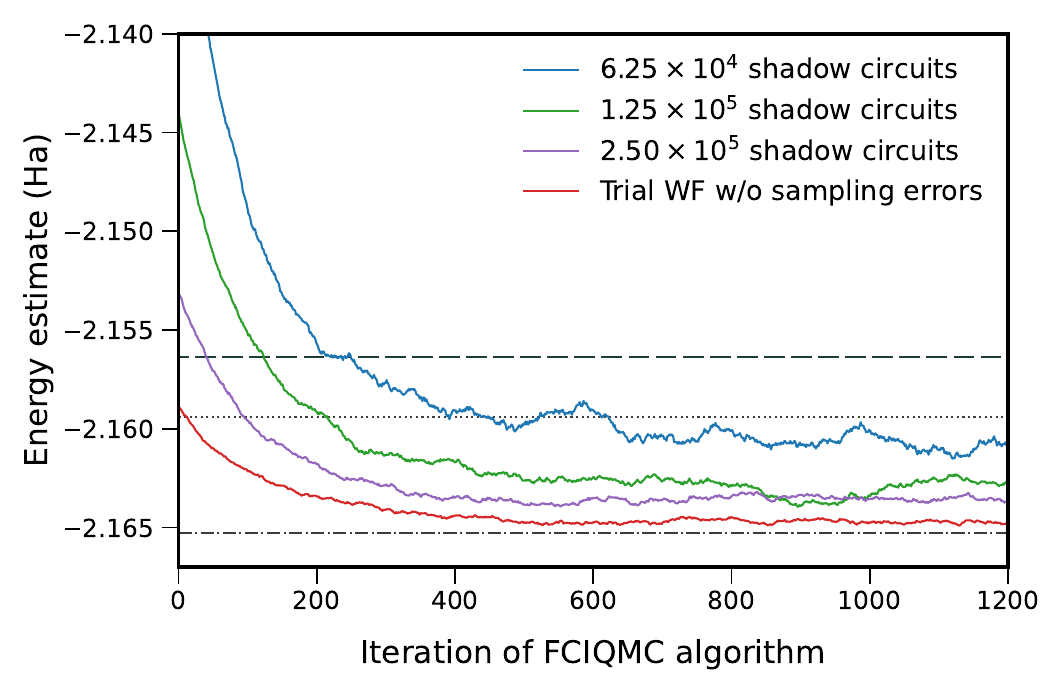}
\caption{Example fixed-node FCIQMC simulations for the linear H$_4$ molecule in a STO-6G basis. Simulations are performed using either the exact LUCJ trial function (red), or using overlaps generated from the classical shadows approach, performed using varying numbers of shadow circuits. The horizontal lines show: the variational energy of the LUCJ wave function, $|\LUCJ\rangle$ (dashed line); the variational of the trial wave function, $|\psiT\rangle$ (dotted line); and the exact ground-state energy of the Hamiltonian (dashed-dotted line).}
\label{fig:simulation}
\end{figure}

\begin{table*}[t]
\begin{center}
{\footnotesize
\begin{tabular}{@{\extracolsep{4pt}}l@{\hskip 5mm}cc@{\hskip 5mm}cc@{\hskip 5mm}cc@{}}
\hline
\hline
 & \multicolumn{2}{c}{1-qubit Cliffords} & \multicolumn{2}{c}{4-qubit Cliffords} & \multicolumn{2}{c}{8-qubit Cliffords} \\
\cline{2-3} \cline{4-5} \cline{6-7}
$\#$ of shadow circuits & error (mHa) & $\%$ error removed & error (mHa) & $\%$ error removed & error (mHa) & $\%$ error removed \\
\hline
  15625  &  16.5481  &  -85.3   &   2.7674  &  69.0   &   1.3223  &   85.2  \\
  31250  &  11.7277  &  -31.3   &   2.3080  &  74.2   &   0.9739  &   89.1  \\
  62500  &   4.7271  &   47.1   &   1.2364  &  86.2   &   0.8747  &   90.2  \\
$1.25 \times 10^5$  &   2.3942  &   73.2   &   0.9404  &  89.5   &   0.7506  &   91.6  \\
$2.5 \times 10^5$  &   1.5718  &   82.4   &   0.6070  &  93.2   &     -     &      -  \\ 
$5 \times 10^5$  &   1.1741  &   86.9   &      -    &     -   &     -     &      -  \\ 
$1 \times 10^6$  &   0.8823  &   90.1   &      -    &     -   &     -     &      -  \\ 
\hline
\hline
\end{tabular}
}
\caption{Numerical results for the fixed-node FCIQMC procedure on H$_4$ in a STO-6G basis, performed using different ensembles, $\mathcal{U}$, and numbers of shadow circuits used to estimate overlaps. For the exact LUCJ trial wave function, $92.1\%$ of error is removed by the fixed-node approximation, compared to $\EvarLUCJ$, which is the energy that would be obtained by VQE.}
\label{tab:h4_fciqmc}
\end{center}
\end{table*}

\begin{figure}[t]
\includegraphics[width=\linewidth]{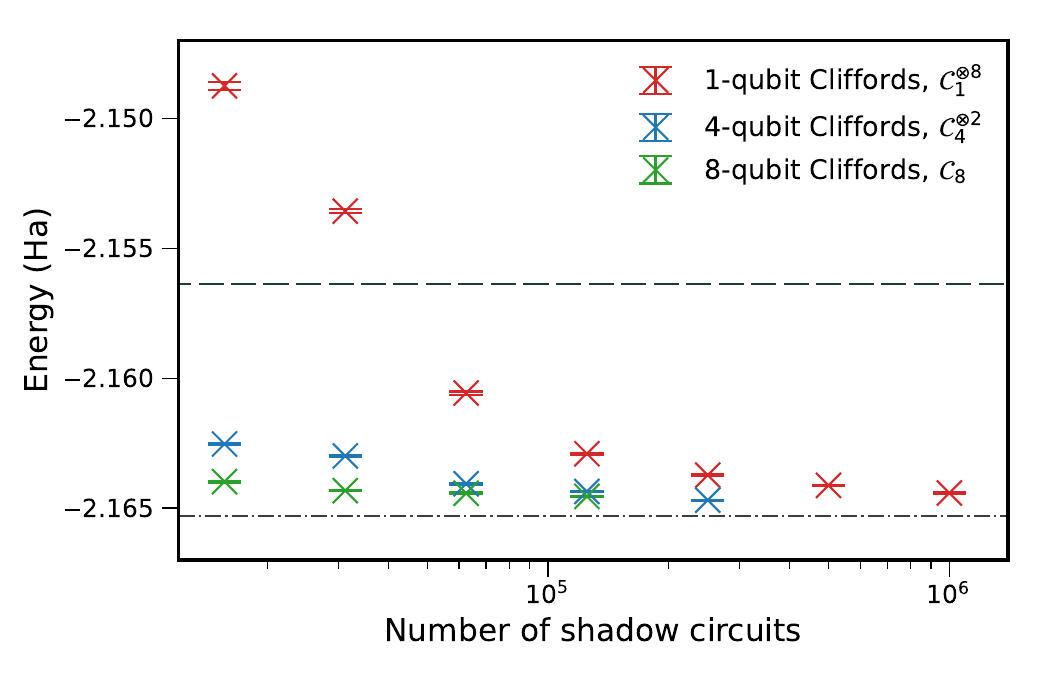}
\caption{Fixed-node energies for the linear H$_4$ molecule, obtained from the fixed-node FCIQMC approach with classical shadows. These energies are obtained by averaging after convergence of the fixed-node FCIQMC simulations. Performing classical shadows using $\mathcal{C}_4^{\otimes 2}$ or $\mathcal{C}_8^{\otimes 1}$ ensembles leads to more rapid convergence than tensor products of 1-qubit Cliffords. The variational LUCJ energy ($\EvarLUCJ$, dashed line) and exact ground-state energy (dashed-dotted line) are shown for comparison.}
\label{fig:h4_fciqmc}
\end{figure}

We first consider application to the linear H$_4$ molecule in a STO-6G basis, using 8 qubits. The exact ground-state energy for this system is $-2.16529$~Ha. The LUCJ ansatz as shown in Fig.~\ref{fig:lattice_and_circuits} was used. Optimizing this LUCJ wave function to minimize the variational energy estimate, we obtain a variational energy ($\EvarLUCJ$) in error by $8.9$~mHa. We then define the trial wave function to be used in fixed-node FCIQMC by the procedure described in Sec.~\ref{sec:real_trial_wf}. This improves upon the LUCJ wave function, giving an error in the variational energy ($\EvarT$) of $5.9$~mHa.

Fig.~\ref{fig:simulation} shows example fixed-node FCIQMC simulations, using a step size $\Delta\tau = 0.005$, and fixed-node parameter $\gamma=0$, corresponding to the standard fixed-node approximation. Simulations are initialized by sampling from the probability distribution proportional to $|\langle D_i | \psiT \rangle|^2$ using the variational Monte Carlo (VMC) algorithm. Results labelled ``Trial WF w/o sampling errors'' show a simulation where overlaps $\langle D_i | \psiT \rangle$ are obtained by direct and exact computation. From this, the true fixed-node energy is obtained as $-2.16459(2)$~Ha. The final fixed-node error is $0.7$~mHa, corresponding to $92.1\%$ of error removed from the variational energy of $|\LUCJ\rangle$, or $88.0\%$ of error removed from the variational energy of $|\psiT\rangle$.

The remaining simulations in Fig.~\ref{fig:simulation} use the classical shadows procedure to estimate the wave function overlaps. Here, Clifford shadows from tensor products of single-qubit Cliffords, $\mathcal{C}_1^{\otimes 8}$, are used for all results. As the number of measurements is increased, the final FCIQMC energy tends towards the true fixed-node energy, as expected. Note that the FCIQMC simulations are initialized such that the energy at iteration $0$ provides an estimate of the variational energy for the trial wave function, $\EvarT$. Even when this initial energy is in significant error, the final energy converged upon is often below $\EvarT$. We also mention again that the energy at iteration $0$ is below $\EvarLUCJ$ for the LUCJ wave function, because the imaginary component of $| \LUCJ \rangle$ is projected away to obtain $ |\psiT \rangle$, as in Eq.~\ref{eq:real_component}, which improves the variational energy.

Fig.~\ref{fig:h4_fciqmc} presents energies calculated by the fixed-node FCIQMC approach using three different ensembles, $\mathcal{U}$, for the classical shadows procedure: tensor products of single-qubit ($\mathcal{C}_1^{\otimes 8}$) and four-qubit ($\mathcal{C}_4^{\otimes 2}$) Clifford circuits, and eight-qubit Clifford circuits ($\mathcal{C}_8$). As expected, the latter two ensembles significantly reduce the number of measurements needed to converge the fixed-node energy, and reduce error at low numbers of measurements. However, the sampling overhead remains very high. To achieve convergence to $0.2$~mHa from the true fixed-node energy, $\sim 6.25 \times 10^4$ measurements were needed with $\mathcal{C}_8$, or $\sim 10^6$ measurements with $\mathcal{C}_1^{\otimes 8}$. Nonetheless, even for the smallest number of measurements considered, the variational energy estimates of both $| \LUCJ \rangle$ and $| \psiT \rangle$ are improved upon.

Table~\ref{tab:h4_fciqmc} gives numerical values for the associated data in Fig.~\ref{fig:h4_fciqmc}, including the percentage of error removed from $\EvarLUCJ$, compared to the exact ground-state energy. With sufficient samples, over $90\%$ of error is removed, and sub-mHa accuracy can be achieved. However, for sufficiently small sample sizes (as seen for $\mathcal{C}_1^{\otimes 8}$ here), energies can be significantly above the variational energy of $| \LUCJ \rangle$.

\subsection{Simulations under circuit-level noise}
\label{sec:noise}

\begin{figure*}[t]
\includegraphics[width=1.0\linewidth]{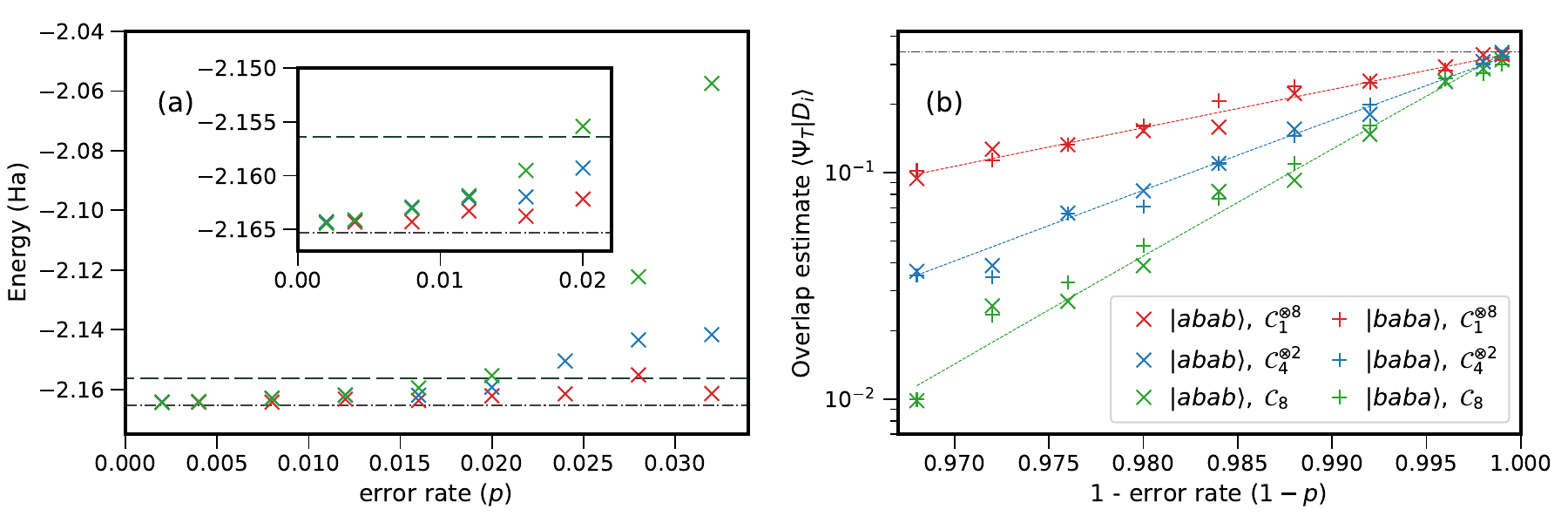}
\caption{Results from simulations under circuit-level depolarizing noise for linear H$_4$, varying the error rate, $p$. (a) Energies calculated from the fixed-node FCIQMC approach, using the Clifford shadows generated under depolarizing noise. Horizontal lines show: the variational LUCJ energy, $\EvarLUCJ$ (dashed line); and the exact ground-state energy (dashed-dotted line). (b) Overlap estimates for two determinant, $| abab \rangle$ and $| baba \rangle$. In the absence of finite sampling errors, these determinants have the same overlap with $| \psiT \rangle$ due to symmetry. Although the overlap estimates vary depending on the error rate and ensemble used, the ratio remains roughly constant. The exact overlap is shown by the dashed-dotted line.}
\label{fig:circuit_level_noise}
\end{figure*}

Next we consider the performance of the method under circuit-level depolarizing noise. The depolarizing channel can be defined
\begin{equation}
    \Delta(\rho) = (1 - p) \rho + \frac{p}{2^n} \mathbbm{1},
\end{equation}
where $n = 1$ for one-qubit gates and $n = 2$ for two-qubit gates, and $p$ is the error rate.

We apply the following simple error model:
\begin{itemize}
    \item a single-qubit depolarizing channel with error rate $p/10$ applied to RX gates,
    \item a two-qubit depolarizing channel with error rate $p$ applied to two-qubit gates.
\end{itemize}
For simplicity we do not apply measurement errors, which will not affect results significantly. We also do not apply errors to RZ gates, which can often be performed virtually \cite{mckay_2017}.

We compile to a native gate set consisting of single qubit gates (RX and RZ) and two-qubit gates (CZ, CPHASE and XY). The LUCJ circuit used is as shown in Fig.~\ref{fig:lattice_and_circuits}, with each Givens rotation decomposed as
\begin{equation}
    G(\theta, \beta) = \mathbb{1} \otimes RZ(-\beta) \circ XY(-\theta) \circ \mathbb{1} \otimes RZ(\beta),
\end{equation}
with
\begin{equation}
\hspace{-3mm}
XY(\theta) = 
\begin{pmatrix}
1 & 0 & 0 & 0\\[0.1em]
0 & \mathrm{cos}(\frac{\theta}{2}) & i \, \mathrm{sin}(\frac{\theta}{2}) & 0\\[0.2em]
0 & i \, \mathrm{sin}(\frac{\theta}{2}) & \mathrm{cos}(\frac{\theta}{2}) & 0\\[0.1em]
0 & 0 & 0 & 1
\end{pmatrix}, \,\,
RZ(\theta) = 
\begin{pmatrix}
e^{-\frac{i\theta}{2}} & 0\\[0.1em]
0 & e^{\frac{i\theta}{2}}
\end{pmatrix}.
\end{equation}
Note that the $XY$ gate is an instance of an fSIM native gate.

To sample and compile random $n$-qubit Cliffords for $n>1$, we follow the procedure outlined in \cite{huggins_2022}. This uses the results by Bravyi and Maslov \cite{bravyi_2021}, where the authors present an algorithm to uniformly generate random Cliffords in a canonical form, and also show that a Clifford in this form can be converted into stages -X-Z-P-CX-CZ-H-CZ-H-P-. The stages -X-Z-P-CX-CZ- will only permute and add a phase to a computational basis state, and so need not be implemented in practice. The remaining stages -H-CZ-H-P- can be implemented using the result of Maslov and Roetteler \cite{maslov_2018}, which shows that an arbitrary CZ circuit followed by qubit reversal can be compiled to a circuit with CNOT depth $2n+2$, using only a linear qubit topology. The effect of the qubit reversal can be accounted for in post-processing. In this way, we sample from an ensemble equivalent to $\mathcal{C}_n$ using circuits with two-qubit gate depth of only $2n+2$. A more detailed explanation is given in Supplementary information of \cite{huggins_2022}. Therefore for the following results, for $\mathcal{U} = \mathcal{C}_4^{\otimes 2}$ and $\mathcal{U} = \mathcal{C}_8$ the measurement operator has CZ depth $10$ and $18$, respectively.

Ref.~\cite{huggins_2022} gives a simple argument to suggest that the QC-QMC approach should have good tolerance to global depolarizing errors. As in Eq.~\ref{eq:overlap_eq_1}, overlaps are estimated by
\begin{equation}
    \langle D_i | \psiT \rangle =  2 \, \textrm{Tr} \big[ \, | 0 \rangle \langle D_i | \, \rho \, \big].
\end{equation}
Therefore, if $\rho$ undergoes a global depolarizing channel with error rate $p$, then since $| 0 \rangle \langle D_i |$ is traceless, one expects that the estimate of $\langle D_i | \psiT \rangle$ will be simply rescaled by a factor $1 - p$, and the same rescaling factor applies to \emph{all} overlaps. Overlaps first enter the fixed-node QMC simulation through ratios of the form $\langle D_i | \psiT \rangle / \langle D_j | \psiT \rangle$, and such rescaling factors due to depolarizing errors should cancel out. Overlaps also enter through the sign of $s_{ij}$ defined in Eq.~\ref{eq:sij}, but again, rescaling due to depolarizing errors will not flip these signs. However, these arguments are only exactly true in the absence of statistical noise. Therefore they are only expected to hold approximately in practice, and only for sufficiently low error rates.

We consider the same linear H$_4$ system studied in Section~\ref{sec:h4}. Results are presented in Fig.~\ref{fig:circuit_level_noise}, and are shown for $\mathcal{C}_1^{\otimes 8}$ (red), $\mathcal{C}_4^{\otimes 2}$ (blue) and $\mathcal{C}_8$ (green) ensembles in the classical shadows procedure. To ensure that results are largely converged with respect to finite sampling error at $p=0$, we perform $5 \times 10^4$ shadow circuits for the latter two ensembles, and $2.5 \times 10^5$ shadow circuits for the former ensemble. For this particular system and ansatz, we find that the method remains accurate for error rates up to around $1\%$. Beyond this, simulations performed with $\mathcal{C}_4^{\otimes 2}$ and $\mathcal{C}_8$ Clifford shadows begin to perform more poorly, and eventually the method fails.

Fig.~\ref{fig:circuit_level_noise}(b) provides insight into this result. Here we plot the overlap estimates $\langle D | \psiT \rangle$ for two determinants with configurations $|abab\rangle$ and $|baba\rangle$. Due to symmetry, the overlap with respect to both determinants will be exactly equal in the absence of errors. We plot these overlaps for all three ensembles considered, taking error rates from $p=0.001$ to $p=0.032$, and for each ensemble fit the data according to $E = a(1-p)^b$. We observe that the ratio of overlaps $\langle D_i | \psiT \rangle / \langle D_j | \psiT \rangle$ is roughly constant. However, as the error rate increases, particularly when sampling from $8$-qubit Clifford circuits, the signal from each overlap tends towards $0$, and the accuracy is eventually limited by the statistical precision due to the classical shadows procedure. This eventually leads to a breakdown in the performance of the approach, as seen in Fig.~\ref{fig:circuit_level_noise}(a).

Despite the eventual breakdown, this tolerance to errors is likely to be valuable in practice. While this tolerance only holds for incoherent errors, coherent errors can often be mitigated via techniques such as randomized compiling (RC) \cite{wallman_2016, hashim_2021}. In general, RC cannot be used to Pauli twirl non-Clifford two-qubit gate layers, and so would either require expressing the circuit in terms of Clifford two-qubit gates, or twirling over an alternative commuting set \cite{kim_2023}. Such an approach would be valuable to test in future experimental studies of the QC-QMC approach. We also note that Ref.~\cite{huang_2024} recently performed a similar study of overlaps and overlaps ratios estimated by Matchgate shadows, where the noise resilience is less clear theoretically. A similar result was found, including when performing experiments on trapped ion processors (and without the use of noise tailoring techniques). We will return again in Sec.~\ref{sec:afqmc} to consider the performance of the QC-QMC methodology in the presence of depolarizing errors, when comparing performance to the AFQMC method.

\subsection{Ferrocene and benzene}
\label{sec:ferrocene_benzene}

Next we consider active spaces with $5$ and $6$ spatial orbitals. Specifically, we consider ferrocene in a $(6\mathrm{e}, 5\mathrm{o})$ active space, and benzene in a $(6\mathrm{e}, 6\mathrm{o})$ active space; see Sec.~\ref{sec:implementation} for details. Note that the fixed-node FCIQMC method was also applied to ferrocene active spaces in Ref.~\cite{blunt_2021}, where it was found that the fixed-node approximation was less successful than for other systems.

For the ferrocene example, the exact ground-state energy is $-1656.02266$~Ha. The variational energy of $| \LUCJ \rangle$ is in error by $6.2$~mHa, while the variational energy of $| \psiT \rangle $ (obtained by projecting away imaginary components of $| \LUCJ \rangle$, as described in Section~\ref{sec:real_trial_wf}) is in error by $4.5$~mHa. Performing the fixed-node FCIQMC method with $| \psiT \rangle $ and using exact overlaps, we obtain a fixed-node energy of $-1656.02024(8)$~Ha, corresponding to an error of $2.42(8)$~mHa. Therefore, the fixed-node approximation removes $46.5\%$ of error compared to the variational energy of $| \psiT \rangle $, or $61.4\%$ compared to $| \LUCJ \rangle$. This is larger than than the fractional improvement of around $36\%$ observed in \cite{blunt_2021} for slightly larger ferrocene active spaces. However, it is below the improvement seen for other systems investigated, which is typically around $70\%$ or more, and so the same trend regarding the difficulty of this ferrocene example is observed. We speculate that the fixed-node approximation is most accurate when performed in a localized basis set; indeed, this method was observed to be less effective when using canonical or split-localized basis sets in \cite{blunt_2021}. Since the orbitals in the $(6\mathrm{e}, 5\mathrm{o})$ basis all have significant $d$-type character on the iron atom, it is not possible to localize the orbitals properly. This is in contrast to the benzene example studied below, where each carbon atom contributes one $p$ orbital, so that the orbital localization procedure is effective.

\begin{figure}[t]
\includegraphics[width=\linewidth]{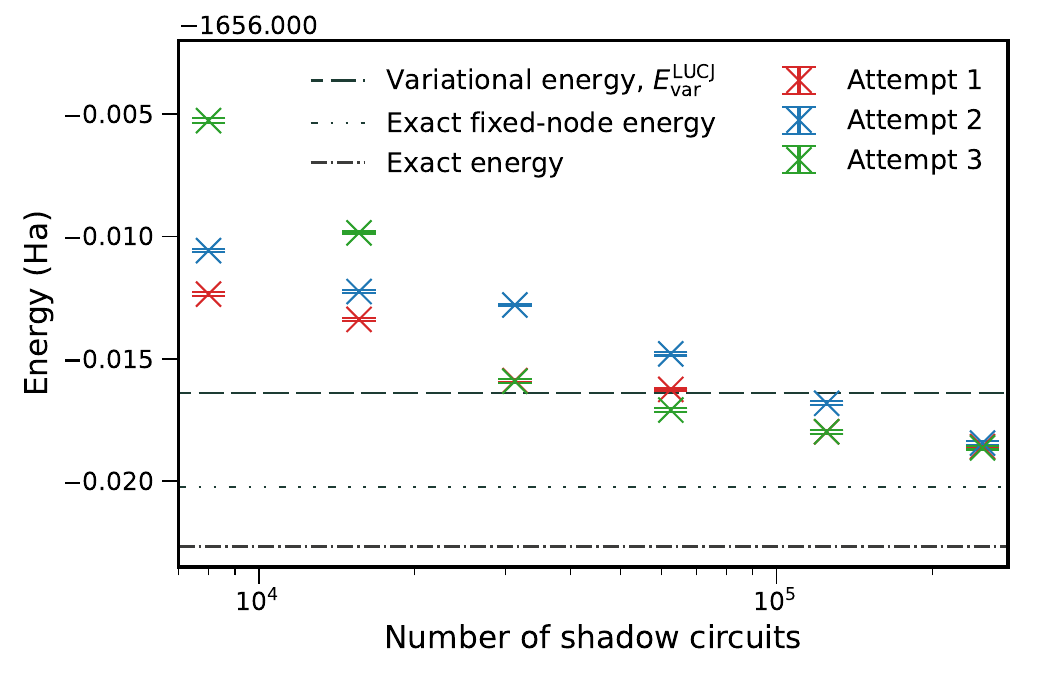}
\caption{Fixed-node energies for the ferrocene molecule in a $(6\mathrm{e}, 5\mathrm{o})$ active space, as a function of the number of shadow circuits performed. Clifford shadows with two partitions ($C_5^{\otimes 2}$) were used. The classical shadows experiment was performed three times with different random number seeds, showing the variance that can occur between runs. Results are not fully converged with $2.5 \times 10^5$ shadow circuits used.}
\label{fig:ferrocene}
\end{figure}

Results for ferrocene with the classical shadows procedure are presented in Fig.~\ref{fig:ferrocene}. Here we perform three separate repeats of the classical shadows experiment, each performing up to $2.5 \times 10^5$ shadow circuits, with a different random number seed for each repeat. These results were obtained using Clifford shadows with two partitions, $\mathcal{C}_5^{\otimes 2}$. Repeating the procedure emphasizes that the statistical error in the overlap estimates naturally leads to some variation in the final fixed-node results, which can be observed here. Despite performing up to $2.5 \times 10^5$ shadow circuits, the estimated fixed-node energy remains above the true fixed-node energy by $1.6$ to $1.8$~mHa for the three repeated experiments, and further sampling would be required for a converged result.

For the benzene example, the exact energy is $-230.69993$~Ha. The variational energy of $| \LUCJ \rangle$ is in error by $28.5$~mHa. We note that the variational optimization of $| \LUCJ \rangle$ was challenging, such that the variational energy may not be optimal for this ansatz and system, however it is sufficient for our purposes here. The variational energy of $| \psiT \rangle $ is in error by $23.5$~mHa. Performing the fixed-node FCIQMC method with $ | \psiT \rangle $, a fixed-node error of $3.1$~mHa is obtained. This corresponds to $86.6\%$ of error removed from the variational energy of $| \psiT \rangle $, or $88.9\%$ removed from the variational energy of $| \LUCJ \rangle$. Therefore, the fixed-node approximation performs significantly better for benzene in this active space, compared to the ferrocene example above. This is also an improvement over the results found in \cite{blunt_2021}, where several acenes were studied in similar $\pi$-type active spaces, and the fixed-node approximation was found to remove around $71\%$ of error, using a different trial wave function.

\begin{figure}[t]
\includegraphics[width=\linewidth]{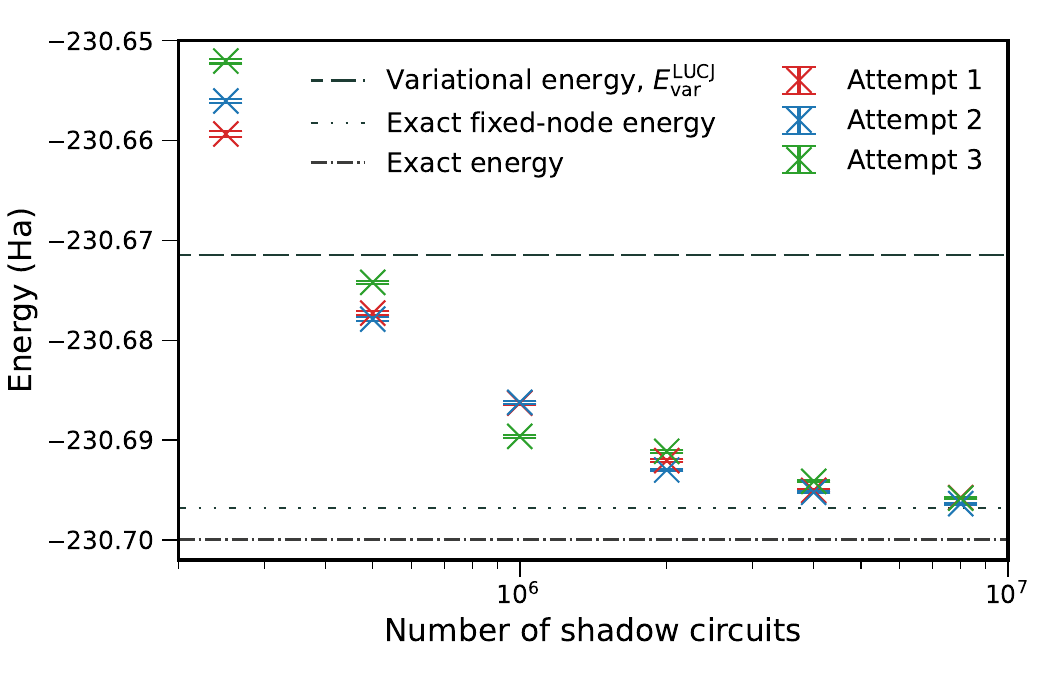}
\caption{Fixed-node energies for the benzene molecule in a $(6\mathrm{e}, 6\mathrm{o})$ active space. Clifford shadows from tensor products of single-qubit Cliffords ($\mathcal{C}_1^{\otimes 12}$) were used. The classical shadows experiment was performed three times with different random number seeds to demonstrate variation between repetitions. Results are eventually converged to $1$ mHa from the exact fixed-node energy in each repetition.}
\label{fig:benzene}
\end{figure}

Results for benzene using the classical shadows approach are presented in Fig.~\ref{fig:benzene}. We again performed three repetitions of the classical shadows experiment, performing up to $8 \times 10^6$ shadow circuits in each. Here, classical shadows used tensor products of single-qubit Cliffords, $\mathcal{C}_1^{\otimes 8}$. With $8 \times 10^6$ shadow circuits, the estimated fixed-node energy is approximately $1$~mHa above the true fixed-node energy in each repetition. Although these results are not fully converged to sub-mHa accuracy, this corresponds to removing $82.2\%$ of error from the variational energy of $ | \psiT \rangle $, a significant improvement. Nonetheless, the sampling cost to achieve this is very high.

In addition to the standard fixed-node approximation, performed with $\gamma = 0$ in the fixed-node Hamiltonian, it is also possible to consider the performance with different values of $\gamma$. For $\gamma = -1$ the exact Hamiltonian is retrieved, while $-1 < \gamma < 0$ is known as the partial-node approximation \cite{kolodrubetz_2012}. Reducing $\gamma$ towards $-1$ is guaranteed to improve the energy of $\Hfn(\gamma)$ towards the exact ground-state energy. However, performing QMC with $\Hfn(\gamma)$ for $-1 < \gamma < 0$ will allow sign violations in the propagation, therefore reintroducing a sign problem, whose severity will grow with decreasing $\gamma$. Nonetheless, it is known that FCIQMC can perform stable sampling in the presence of a sign problem, provided that the walker population is sufficiently large \cite{spencer_2012},which we choose to be the case here. Therefore, sampling with FCIQMC allows us to consider such as regime.

Moreover, because $\Efn(\gamma)$ is a concave function \cite{beccaria_2001}, any linear extrapolation of $E(\gamma_1)$ and $E(\gamma_2)$ to $\gamma=-1$, for $\gamma_1, \gamma_2 > -1$, is guaranteed to give both a variational and improved ground-state energy estimate (provided that statistical errors are controlled). Since we can take both $\gamma_1 \ge 0$ and $\gamma_2 \ge 0$, no sign problem need be introduced to perform this approach, and a small walker population can be used.

\begin{figure}[t]
\includegraphics[width=\linewidth]{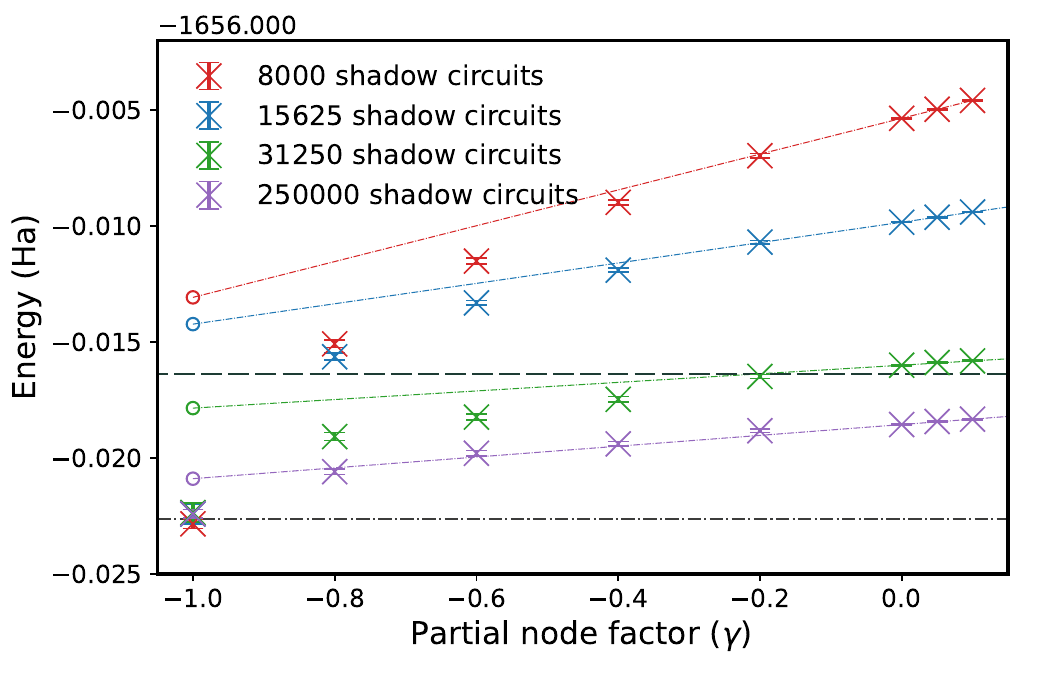}
\caption{Partial-node approximation using the classical shadows approach for a ferrocene example. Horizontal lines show: the variational LUCJ energy, $\EvarLUCJ$ (dashed line); and the exact ground-state energy (dashed-dotted line). Results at $\gamma=-1$ corresponds to the exact Hamiltonian, while $\gamma=0$ is the standard fixed-node approximation. With FCIQMC sampling, it is possible to perform calculations for $\gamma < 0$, although this reintroduces a sign problem. Also plotted are extrapolations to the exact $\gamma=-1$ energy, using $\Efn(0.0)$ and $\Efn(0.1)$. Because $\Efn(\gamma)$ is a concave function, this gives an improved variational estimate of the energy.}
\label{fig:partial_node}
\end{figure}

Results are presented in Fig.~\ref{fig:partial_node} to demonstrate both of these approaches. We consider the ferrocene example from above which proved challenging. As an example, we choose the same set of classical shadows used for ``Attempt 3'' in Fig.~\ref{fig:ferrocene}, and consider four different numbers of samples. As expected, energies systematically converge to the exact ground-state energy as $\gamma$ is reduced, with the exact result obtained at $\gamma = -1$. So, partially lifting the fixed-node approximation provides one path to reduce dependence on the number of circuits in the classical shadows procedure. However, we see that even with $\gamma=-0.8$, when performing 15,625 shadow circuits, the calculated energy is still above the variational energy of $| \LUCJ \rangle$. Moreover, while FCIQMC can perform stable sampling in the presence of a sign problem, the walker population to achieve this will scale exponentially with the size of the active space in general. Therefore, we expect it likely that this approach will ultimately become limited in its utility for non-trivial system sizes.

\begin{figure*}[t]
\includegraphics[width=1.0\linewidth]{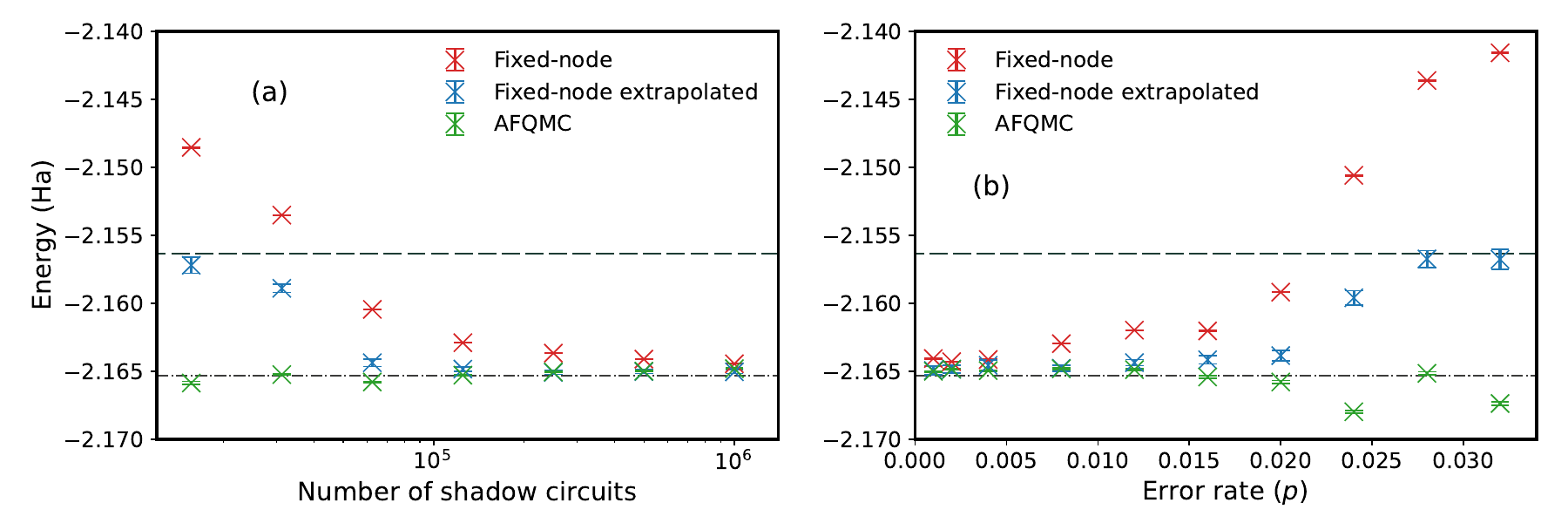}
\caption{Comparison of the fixed-node approximation and the phaseless AFQMC approximation for the linear H$_4$ molecule. Horizontal lines show: the variational LUCJ energy, $\EvarLUCJ$ (dashed line); and the exact ground-state energy (dashed-dotted line). In (a) the circuit is performed without errors, varying the number of shadow circuits, and using shadows from tensor products of single-qubit Cliffords ($\mathcal{C}_1^{\otimes 8}$). In (b) depolarizing errors of varying size are applied, while using partitioned classical shadows with two partitions ($\mathcal{C}_4^{\otimes 2}$), and 50,000 circuits performed. ``Fixed-node'' results are obtained with $\gamma=0$, while ``fixed-node extrapolated'' results are obtained by linear extrapolation of energy estimates with $\gamma=0.0$ and $\gamma=0.1$ to $\gamma=-1$.}
\label{fig:h4_with_afqmc}
\end{figure*}

Also in Fig.~\ref{fig:partial_node}, we plot linear fits using $\Efn(\gamma)$ with $\gamma=0.0$ and $\gamma=0.1$ to obtain an improved estimate at $\gamma = -1$. Unlike the partial-node method, this approach is scalable to large problem sizes. It can be seen that the estimate obtained by this linear extrapolation is often a significant improvement over those obtained from $\Efn(\gamma=0)$, and somewhat reduces dependence on the number of classical shadows used. Even in the limit where the exact overlaps $\langle D_i | \psiT \rangle$ are used, this can significantly reduce the remaining fixed-node error. One potential drawback of this approach is that obtaining an extrapolation of sufficient quality may require small statistical error bars on both estimates of $\Efn$ used. In general this can require performing many iterations in the FCIQMC method, which of course could become a significant cost. A second drawback, as for all approaches based on extrapolations of the energy, is that such extrapolations may be challenging to extend to estimation of other properties.

We will present some further examples of this extrapolation-based approach in the next section, where we return to the H$_4$ system and perform a comparison with AFQMC.

\subsection{Comparison to AFQMC}
\label{sec:afqmc}

We conclude by comparing the accuracy and tolerance to errors in the QC-QMC approach when performed with either the fixed-node approximation or phaseless AFQMC. This also provides a chance to compare the two methods. Previous studies have performed comparison of phaseless AFQMC and diffusion Monte Carlo \cite{malone_2016}, which uses the real-space fixed node approximation \cite{foulkes_2001}, but the determinant-space fixed-node approximation used here is very different in nature. Of course, the accuracy of both methods depends on the system and trial wave function used, and we consider just one example here.

Results are presented in Fig.~\ref{fig:h4_with_afqmc} for the linear H$_4$ system considered in Sections~\ref{sec:h4} and \ref{sec:noise}. Fig.~\ref{fig:h4_with_afqmc}(a) considers results in the absence of depolarizing errors, with unpartitioned classical shadows (sampling from $\mathcal{C}_1^{\otimes 8}$), using the same classical shadows obtained for results in Section~\ref{sec:h4}. Fig.~\ref{fig:h4_with_afqmc}(b) considers results with depolarizing errors and using partitioned classical shadows (sampling from $\mathcal{C}_4^{\otimes 2}$), using the same classical shadows obtained for results in Section~\ref{sec:noise}. In the comparison between the fixed-node and phaseless approximations, we also include the extrapolated fixed-node energies, as previously demonstrated for ferrocene in Fig.~\ref{fig:partial_node}. Here we take the same approach, using energies $\Efn$ obtained with $\gamma = 0.0$ and $\gamma = 0.1$ to perform the extrapolation.

First considering Fig.~\ref{fig:h4_with_afqmc}(a), where the number of shadow circuits is varied, we see that AFQMC is relatively insensitive to the quality of the trial wave function. Performing $10^6$ shadow circuits, the AFQMC energy is in error by $0.49(4)$~mHa, compared to $0.88(2)$ using the fixed-node method, suggesting that, even when overlap estimates are well converged, the phaseless approximation is more accurate. In the low-sampling limit, when performing $1.5625 \times 10^4$ shadow circuits, the fixed-node approximation is in error by $16.5(1)$~mHa, compared to $-0.6(1)$~mHa with phaseless AFQMC. Therefore, while the phaseless approximation becomes non-variational, it is significantly more accurate in the regime where overlaps are poorly estimated. For comparison, when performing $1.5625 \times 10^4$ and $10^6$ shadow circuits, the extrapolated fixed-node approximation gives energies in error by $8.1(5)$~mHa and $0.25(13)$~mHa, respectively. In the former case, around $50\%$ of error is removed from the fixed-node approximation, while the latter is the most accurate among the three approaches, and remains rigorously variational. However, in general the phaseless approximation is the most accurate when using fewer classical shadows.

Next we consider Fig.~\ref{fig:h4_with_afqmc}(b), where the number of classical shadows is fixed and the depolarizing error rate is varied. Here, with an error rate of $p=10^{-3}$, the fixed-node, extrapolated fixed-node and AFQMC methods give energy estimates in error by $1.23(3)$~mHa, $0.36(31)$~mHa and $0.29(4)$~mHa, respectively. In contrast, at the high error rate of $p=0.032$, the energies are in error by $23.6(2)$~mHa, $8.5(7)$~mHa and $-2.1(1)$~mHa, respectively. Again it can be seen that phaseless AFQMC has the best tolerance to errors among the three approaches. However, the extrapolated fixed-node approach removes over $50\%$ of the fixed-node error while remaining rigorously variational, and so offers and interesting improvement over the traditional fixed-node approximation.

The results found here are perhaps not surprising; in the case of weakly-correlated systems, AFQMC is known to give high-accuracy solutions, even with very simple trial wave functions. For example, in weakly-correlated systems, it has been shown that AFQMC with a single-determinant trial wave function typically gives accuracy between that of coupled cluster with singles and doubles (CCSD) and CCSD with perturbative triples (CCSD(T)) \cite{lee_2022_2}. Using configuration interaction singles and doubles (CISD) trial states, the accuracy can be better than CCSD(T) \cite{mahajan_2024}. Therefore, even when the trial state is in significant error, AFQMC can give extremely accurate results. In contrast, the fixed-node approximation requires sophisticated trial wave functions, and each overlap $\langle D_i | \psiT \rangle$ is required to be non-zero for the method to be applicable. The potential benefit of the fixed-node methodology in the classical context is that a larger class of wave functions can be efficiently used. Similarly in the context of QC-QMC, the fixed-node approach avoids the exponential post-processing step when using Clifford shadows to estimate overlaps. However, in the case where a common wave function can equally be used in either method, we expect the AFQMC method to be more accurate in general, particularly when using low numbers of samples such that the trial wave function is poor.

\section{Discussion}
\label{sec:discussion}

In this work we have performed a numerical study of the QC-QMC methodology in combination with fixed-node FCIQMC, using Clifford shadows to construct the required overlap estimates. A potential benefit of this approach is that it avoids the exponential post-processing step that occurs when performing QC-AFQMC with Clifford shadows. This offers an alternative approach to resolve this exponential post-processing step, compared to the use of Matchgate shadows \cite{wan_2023, kiser_2024, huang_2024}.

We performed this study using the local unitary cluster Jastrow (LUCJ) ansatz, and investigated the performance of the method under depolarizing errors, demonstrating its tolerance to noise. We also considered extensions to the standard fixed-node approximation, showing that extrapolations are an effective way to improve the accuracy of the method, and can reduce the dependence on the number of shadow circuits performed. Lastly, we performed a preliminary comparison to AFQMC for the linear H$_4$ example. This suggested that, if a common trial wave function can be used in either method, then the phaseless AFQMC method has better tolerance to errors, although extrapolations of the fixed-node energy reduce this discrepancy.

This work focused on converging the energy within an active space. A separate but important question regards the estimation of virtual correlation energy. Ref.~\cite{huggins_2022} achieved this by performing AFQMC in a larger space with an appropriate trial wave function, such that non-trivial overlap estimation is only required within the active space. This approach would be challenging to extend to the fixed-node method, which requires each determinant in the space sampled to have a non-zero overlap with the trial wave function used. Moreover, the extent to which the fixed-node approximation can be effective in capturing dynamical correlation is an open question. A separate strategy would be to use the approach of Refs.~\cite{mahajan_2019, blunt_2020}; these show that the (strongly-contracted) NEVPT2 energy \cite{angeli_2001} can be calculated using only overlaps within the active space by performing VMC sampling, and so would be amenable to the QC-QMC approach with classical shadows. More generally, this offers an interesting approach to perform NEVPT2 beyond active space VQE calculations.

Despite the avoidance of an exponential post-processing step, alternative approaches need to be identified to reduce the cost of the approach. Even for the small active spaces considered here, the number of classical shadows required is very high. For example, converging to within $1$ mHa of the expected fixed-node energy for linear H$_4$ required $1.25 \times 10^5$ Clifford shadows (using two Clifford partitions). In this paper we have not focused on the time requirement to construct each overlap estimate on-the-fly (this has been a focus of recent QC-AFQMC studies \cite{huang_2024, kiser_2024}), but this will be a limiting factor for large active spaces. There are likely ways to significantly reduce this, for example by precomputing overlaps within a subspace where much of the spectral weight of the eigenstate lies, in the same spirit as semi-stochastic modifications to FCIQMC \cite{petruzielo_2012, blunt_2015}. A more fundamental issue has been raised in \cite{mazzola_2022}, and also first discussed in \cite{huggins_2022}, that although overlaps can be calculated to additive error, relative errors in overlaps and corresponding local energies will grow exponentially with system size in the worst case. Despite this, the benefits of the QC-QMC approach, including noise resilience, shallow circuits suitable for near-term devices, and the potential to expand the set of trial wave functions applicable in QMC, are indeed promising. We hope that future developments will help to resolve these issues and realize this promise.

\section{Acknowledgements}

We thank Earl T. Campbell for valuable comments on this manuscript. This work was supported by Innovate UK via the Quantum Commercialisation programme of the Industrial Strategy Challenge Fund (ISCF) [project No. 10001505].

\bibliography{main}

\end{document}